\documentclass[conference]{IEEEtran} 
\makeatletter 
\def\ps@headings{%
\def\@oddhead{\mbox{}\scriptsize\rightmark \hfil \thepage}%
\def\@evenhead{\scriptsize\thepage \hfil \leftmark\mbox{}}%
\def\@oddfoot{}%
\def\@evenfoot{}} 
\makeatother 
\ifCLASSINFOpdf
  \usepackage[pdftex]{graphicx}
    \graphicspath{{./image/}{./jpeg/}}
    \DeclareGraphicsExtensions{.pdf,.jpeg,.png}
\else
   \usepackage[dvips]{graphicx}
   \graphicspath{{./image/}}
   \DeclareGraphicsExtensions{.eps}
\fi
\usepackage{subfig}
\usepackage{balance}

\begin{document}

\title{Fast and energy-efficient technique for jammed region mapping in wireless sensor networks}

\author{
\IEEEauthorblockN{Nabila Rahman}
\and
\IEEEauthorblockN{Matthew Wright}
\IEEEauthorblockA{Department of Computer Science and Engineering\\
The University of Texas at Arlington\\
\{nabila.rahman@mavs., mwright@, dliu@\}uta.edu
}
\and
\IEEEauthorblockN{Donggang Liu}
}

\maketitle

\begin{abstract}
Wireless sensor networks (WSNs) have great practical importance for surveillance systems to perform monitoring by acquiring and sending information on any intrusion in a secured area. Requirement of very little human intervention is one of the most desirable features of WSNs, thus making it a cheaper and safer alternative for securing large areas such as international borders. Jamming attacks in WSNs can be applied to disrupt communications among the sensor nodes in the network. Since it is difficult to prevent jamming attacks, detection and mapping out the jammed regions is critical to overcome this problem. In a security monitoring scenario, the network operators will be able to take proper measures against jamming once the jammed regions in the network are known to them. It is also desirable to keep the interactions of the sensor nodes in the network minimal, as they are low powered devices and need to conserve their resources. In this paper we propose a light-weight technique for faster mapping of the jammed regions. We minimize the load on the sensors by removing the actual responsibility of mapping from the network to the central base station (BS). After a few nodes report to the BS, it carries out the task of mapping of the jammed regions in the network. We use our simulation results to compare our proposed system with the existing techniques and also to measure the performance of our system. Our results show that the jammed regions in a network can be mapped from fewer nodes reporting to the base station.  
\end{abstract}

\section{Introduction}
\label{intro}

A wireless sensor network (WSN) typically consists of a large number of
autonomous devices with limited battery power and memory. Since cheap
commodity devices can be used as sensor nodes, it is possible to do
large scale deployment of sensor networks~\cite{WSN, WSN2}. They are
ideal for monitoring to detect intruders physically entering a secured
or otherwise important region~\cite{he04energy}. Thus, sensor networks
have a wide variety of applications in security monitoring, such as
protecting water supplies, chemical plants and nuclear power plants and
in border security and battle field surveillance.

\begin{figure}[hpb]
  \begin{center}
    \includegraphics[width = 2.75in]{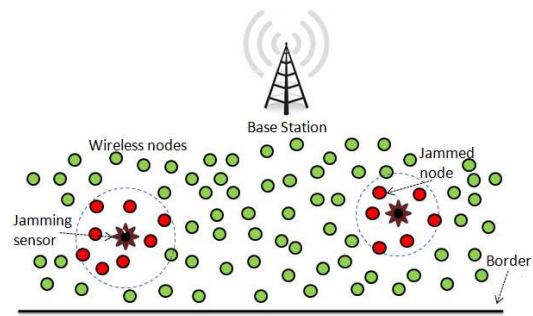}
   \end{center}
  \caption{Wireless sensor network for border monitoring}
  \label{fig:border}
\end{figure}

Typically, WSNs are left unattended for efficient, low-cost
monitoring. Thus, they are deployed without any physical
protection. Also, the nodes in a WSN communicate with each other through
a shared wireless medium. These two features make WSNs particularly
vulnerable to a variety of attacks~\cite{ATT0, ATT1, ATT2}.  Jamming
\cite{FEASJAM, ATTJAM} is a particularly effective attack against
WSNs. An intruder can easily place jamming devices in different parts of
the network to cause radio interference and thus disrupt communications
among the sensor nodes that are in close proximity to the jamming
devices. Fig.~\ref{fig:border} shows a border security deployment
scenario in which the red sensors have been jammed by the jamming sensors present in the network.

A jamming attack effectively creates a denial of service condition in
the network. This a major problem in security monitoring applications,
in which the lack of sensor communication means that an intruder can
physically enter jammed regions without the threat of being
detected. For example, in a border security setting, a path may be
constructed with jamming devices that allows the intruder to cross back
and forth across the border, completely bypassing the security
perimeter. In this case, denial of service in the network leads to a
major breach of physical security. It is critical in these
monitoring applications for the base station to learn about and map out the
jammed regions quickly and accurately so as to know where
physical security may be threatened and where it may be necessary to
increase other security measures like guard patrols or surveillance
flights. 

Wood et al. propose JAM~\cite{JAM}, a jammed area mapping technique,
that relies on the ability of the nodes to perform a detailed mapping
of the jammed region locally. JAM is very effective at mapping out the
jammed region. However, it is also a very complex protocol with high
message and storage overheads at the nodes, due to its fully
decentralized nature. It requires a lot of interaction among the nodes
surrounding the jammed regions to estimate the region and correctly put
the jammed nodes into groups. In settings such as the border security
scenario shown in Fig.~\ref{fig:border}, it is vital to protect the
sensor network and to detect the intruder as early as possible while
keeping communication overhead low to save sensors' battery lifetime.

\noindent {\bf Contributions.} In this paper, we present a model for
studying the jammed sensor mapping problem (\S\ref{model}). We are the
first to point out that mapping jammed nodes should be done {\em
  quickly} and {\em efficiently}, not just accurately. This observation
suggests a different set of design choices than those made in JAM
(briefly described in \S\ref{rel}).

In particular, instead of mapping being performed locally by the sensor
nodes, we leverage the powerful base station to gather information from
the network and calculate where the jammed regions are. In this protocol
(\S\ref{mapprot}), rather than an exact mapping, we only aim to get an
approximation of the jammed area computed by the central base
station. We apply k-means clustering to accurately separate out multiple
jammed regions. We then invoke a method based on convex-hull finding
algorithms to find the centers for these regions and then apply
iterative adjustment to accurately locate and determine the size of the
region. This approach relieves the sensors surrouding a jammed area from
the communication overhead and power consumption of calculating the
jammed regions.


We developed a simulator (\S\ref{sim}) to evaluate our system and
compare it with JAM in terms of effective mapping and communication
overhead. Our results demonstrate that the proposed
protocol performs faster mapping, thereby saving substantial message overhead compared with JAM, and it
provides reasonable mapping accuracy. We also experiment with the
trade-off between the communication overhead of the sensor nodes and the
mapping performance of the system. Finally we run experiments on real sensor notes to see the performance of our jammed region mapping technique (\S\ref{exp}).

%

\section{Related Work}
\label{rel}

Communication in the WSN in the presence of jamming has been investigated previously. Wormhole-based anti-jamming techniques and timing channels are discussed in ~\cite{PERF, cagalj07, Xu2008Antijamming}.Various spread-spectrum communication techniques are used to defend jamming in the wireless networks~\cite{ATTJAM, ATTJAM2, Xu04channelsurfing}. Since, large scale deployment of WSNs mostly have cheap commodity devices as sensor nodes, it is unlikely that they will possess the design complexity to perform spread spectrum techniques, and are more likely to use a single frequency. Some works involve detection of radio interference~\cite{RID_RADIO} in WSNs. The packet delivery ratio (PDR) and the measurement of signal strength can be used to detect jamming in WSNs~\cite{FEASJAM}. Jamming detection by monitoring channel utilization is discussed in JAM \cite{JAM}, where the sensors decide they are jammed when their channel utility is below a certain threshold.

JAM \cite{JAM} is a complete protocol for the jammed area mapping after detection. In this protocol a node that discovers that it has been jammed broadcasts {\em jammed} messages to its neighbors informing them that it has been jammed. The neighboring nodes that are farther in the distance from the jamming effect but located near the boundary of the jammed region will be able to receive these messages and initiate the JAM protocol. By this protocol, the boundary nodes map out the jammed area by exchanging messages among themselves regarding the jammed nodes.

In our protocol, the central BS performs the jammed region mapping.

\section{Model}
\label{model}

In this section we present the basic characteristics of the network and intruder models that we are going to use for our protocol.


\subsection{Network model}
We study our system in a homogeneous network model in which all the nodes are stationary, location aware, and also have roughly the same sensing capabilities. These sensor nodes possess limited power and utilize wireless channels to communicate with other nodes within their signal range.

The sensor nodes close to a jamming device are unable to receive any messages from their neighbors since the channel is jammed  and therefore unable to communicate with any of their neighbors. However, the nodes that are on the edge of a jammed region are able to send messages to their un-jammed neighbors outside the jamming range and send notification messages to them once they detect jamming among some of their neighboring nodes. These nodes are called the boundary nodes.

The base station (BS) is a distinguished component with much more computational power and communication resources than the sensor nodes in the network. The BS is also aware of the network topology and the location of the sensor nodes in the network~\cite{location1, location2}. 

During the deployment phase, flooding messages are sent by the BS to construct a spanning tree rooted at the BS; applying breadth first search. All the nodes within the range of the BS set the BS as their parent and rebroadcast this message to their neighboring nodes. In this way, each node has information about it's predecessor nodes and the minimum distance to the BS, which is used for routing by applying the standard routing algorithms and common pratices~\cite{route1, Schurgers01energyefficient}.

An example network is shown in Fig.~\ref{fig:node}.

\subsection{Intruder model}
In a traditional wireless communication system, a jammer launches jamming attacks on the physical and data link layers of the WSNs with the goal of preventing reception of communications at the receiver end using as little power as possible.

In this paper, we consider an intruder who can deploy jamming devices that act as constant jammers. The intruder can place these malicious nodes or jamming devices at arbitrary locations in the network for the purpose of creating a DOS condition or an unmonitored path through the network. 

We assume that the jamming devices have similar hardware capabilities as the sensor nodes in the network, and that they act as constant jammers. Thus they disrupt any communication in the surrounding area by emitting continuous radio signals. They can be implemented using regular wireless devices that continuously sending out random bits to the channel without following any MAC-layer protocol~\cite{FEASJAM} and thus effectively being able to block the legitimate traffic sources from using the channel. The range of these jamming devices may vary, and their range is not known to the sensor nodes or the BS. 

For example, in Fig.~\ref{fig:events}, we present a scenario in which there are five jamming sensors in the network. The red nodes are the sensor nodes that are jammed and are unable to communicate with the unjammed nodes marked in green.

\section{Mapping protocol}
\label{mapprot}

\begin{figure}
  \begin{center}
    \subfloat[WSN]{\label{fig:node}\fbox{\includegraphics[width = 1.5in]{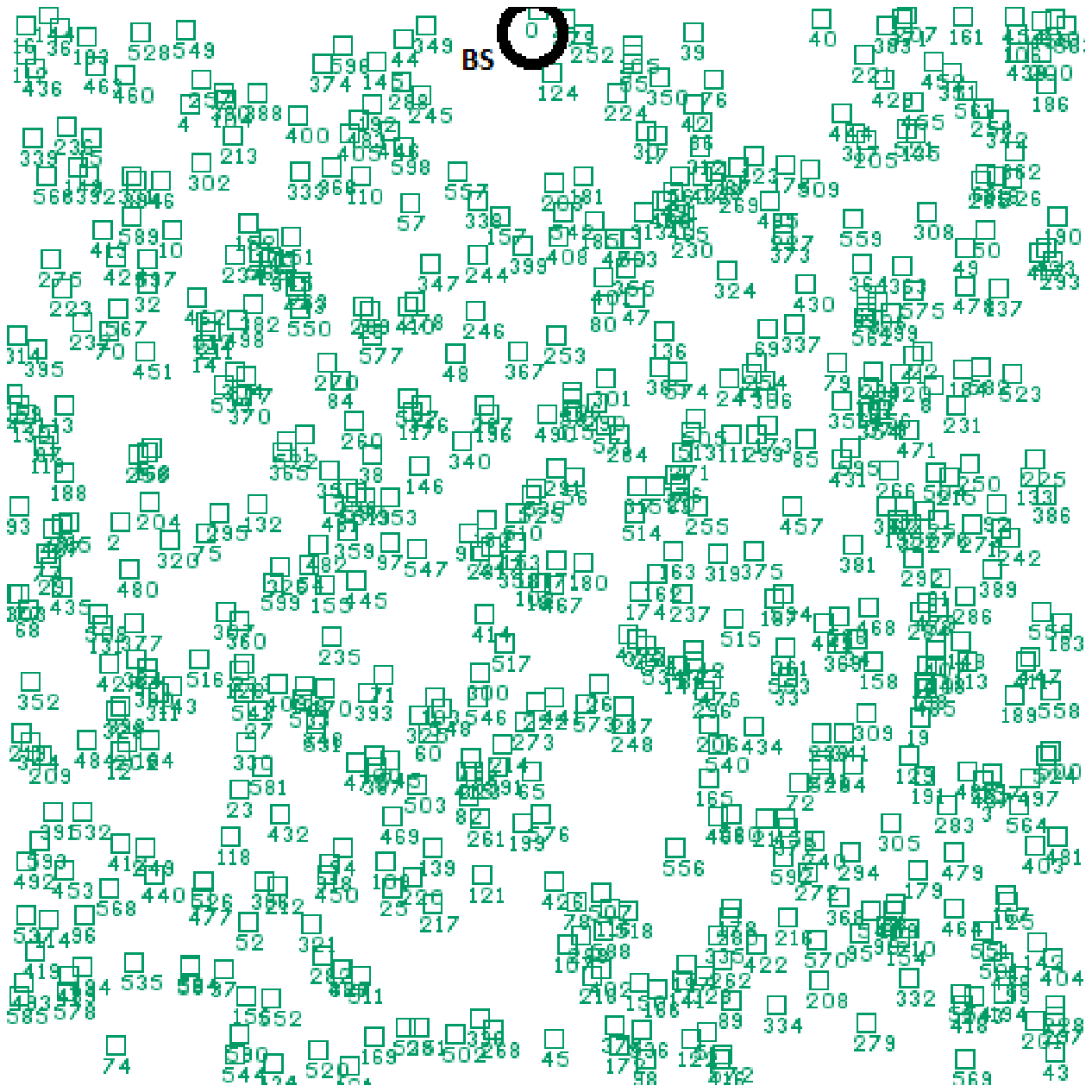}}}
    \hspace{.1in}    
    \subfloat[Jamming in WSN]{\label{fig:events}\fbox{\includegraphics[width = 1.5in]{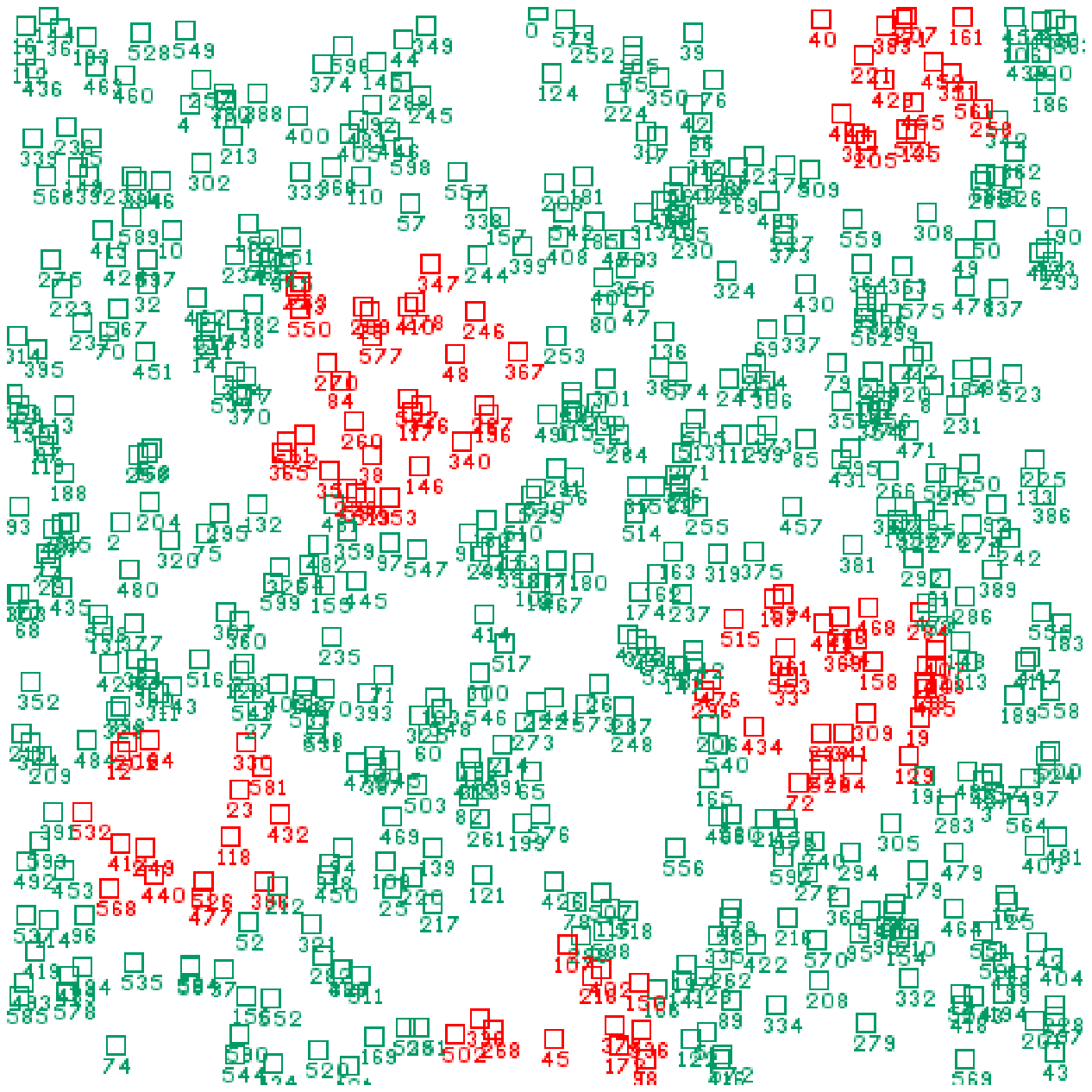}}}
  \end{center}
   \begin{center}
    \subfloat[Reporter nodes]{\label{fig:rep}\fbox{\includegraphics[width = 1.5in]{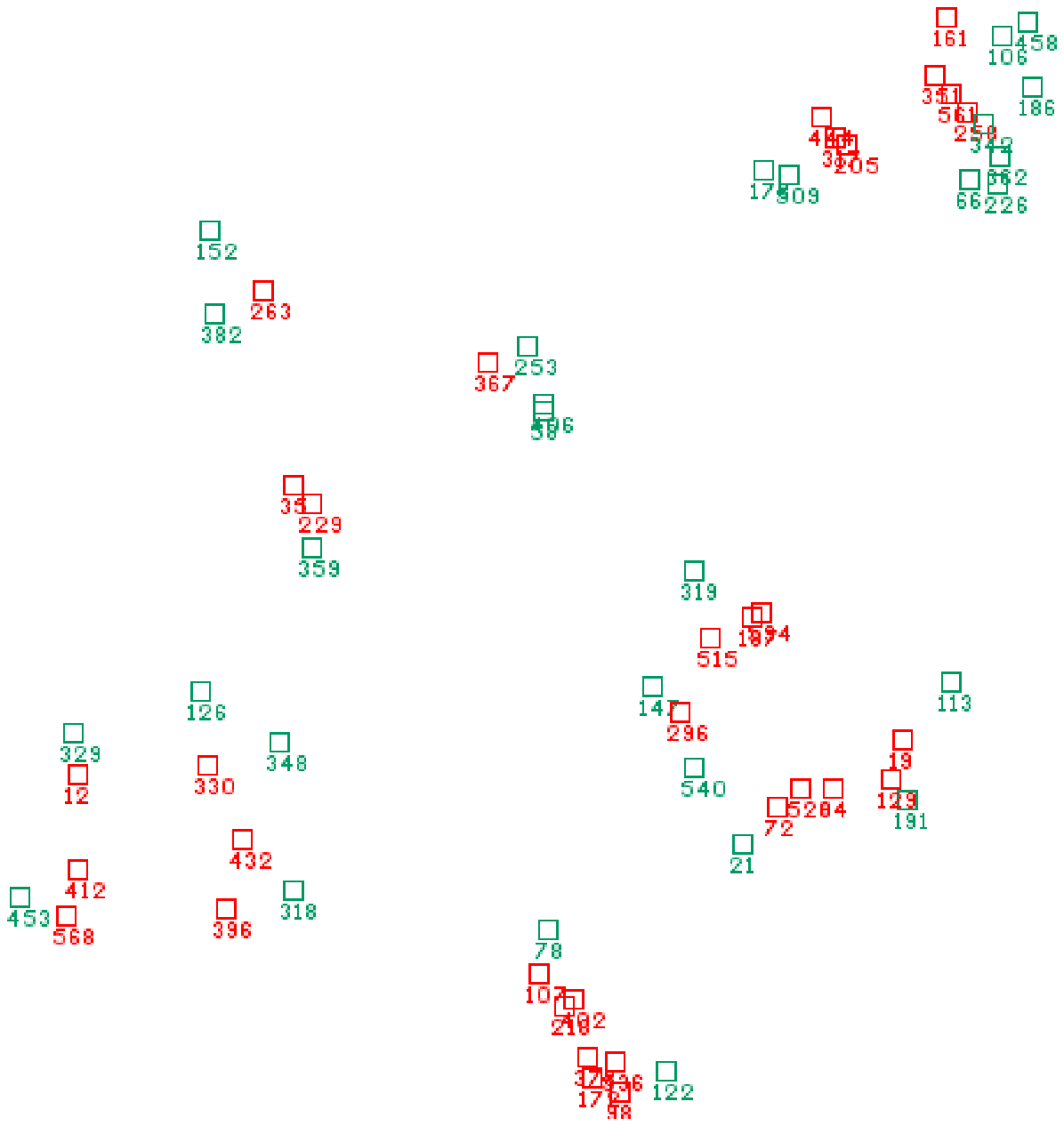}}}
    \hspace{.1in}    
    \subfloat[Finding jammed regions]{\label{fig:clust}\fbox{\includegraphics[width = 1.5in]{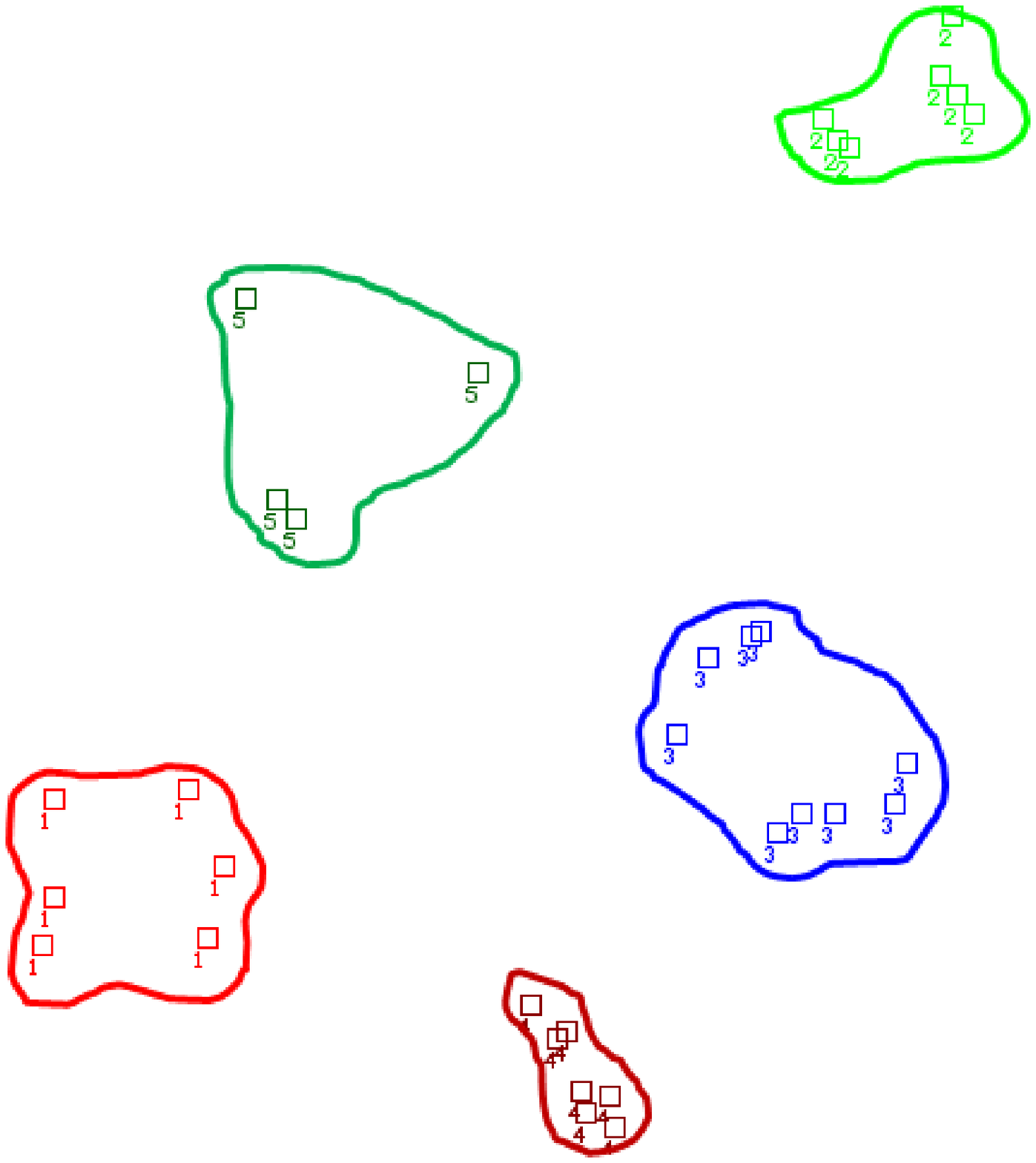}}}
  \end{center}
  \begin{center}
    \subfloat[Convex hull of the regions]{\label{fig:hull}\fbox{\includegraphics[width = 1.5in]{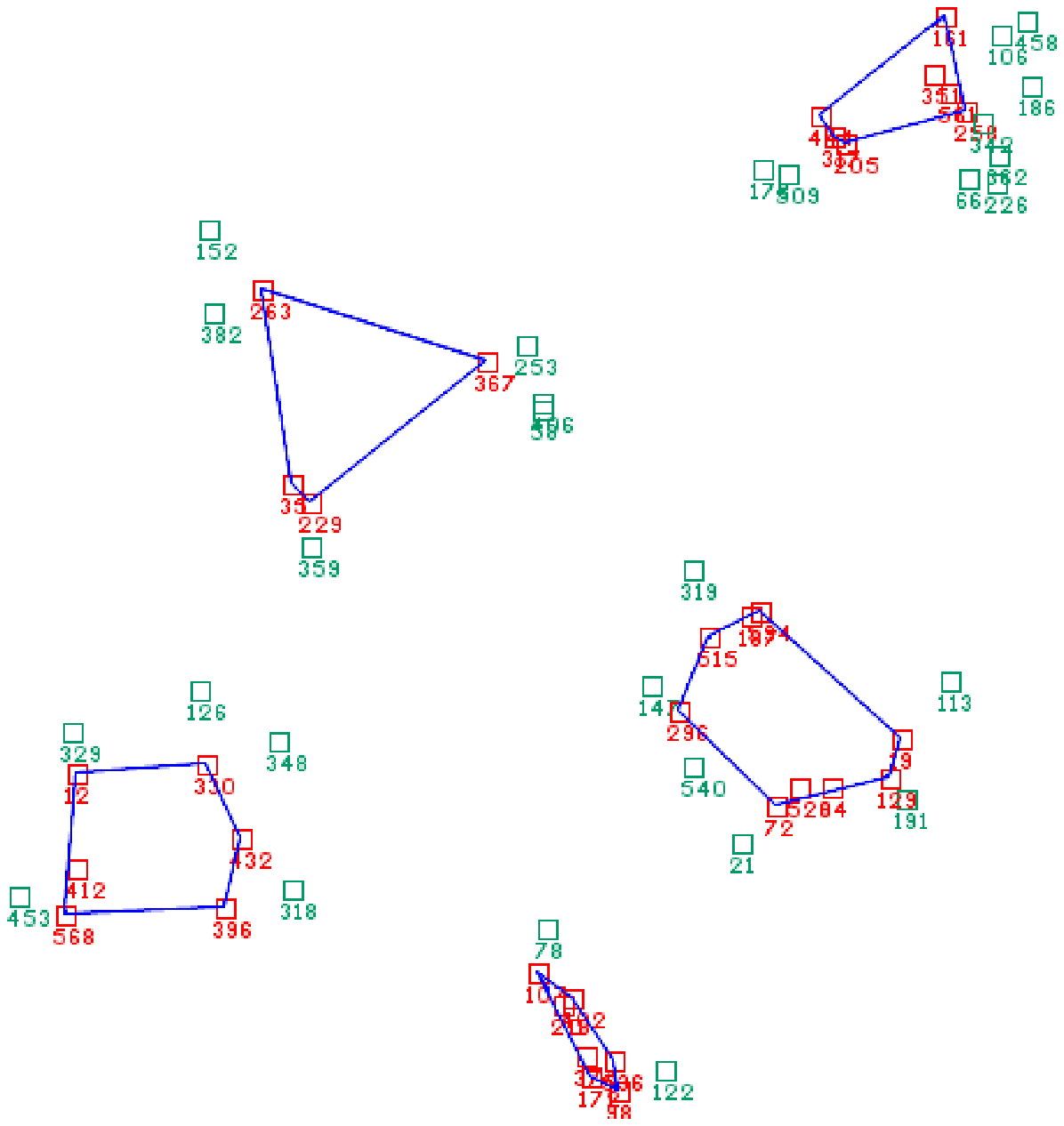}}}
    \hspace{.1in}    
    \subfloat[Mapped area]{\label{fig:area}\fbox{\includegraphics[width = 1.5in]{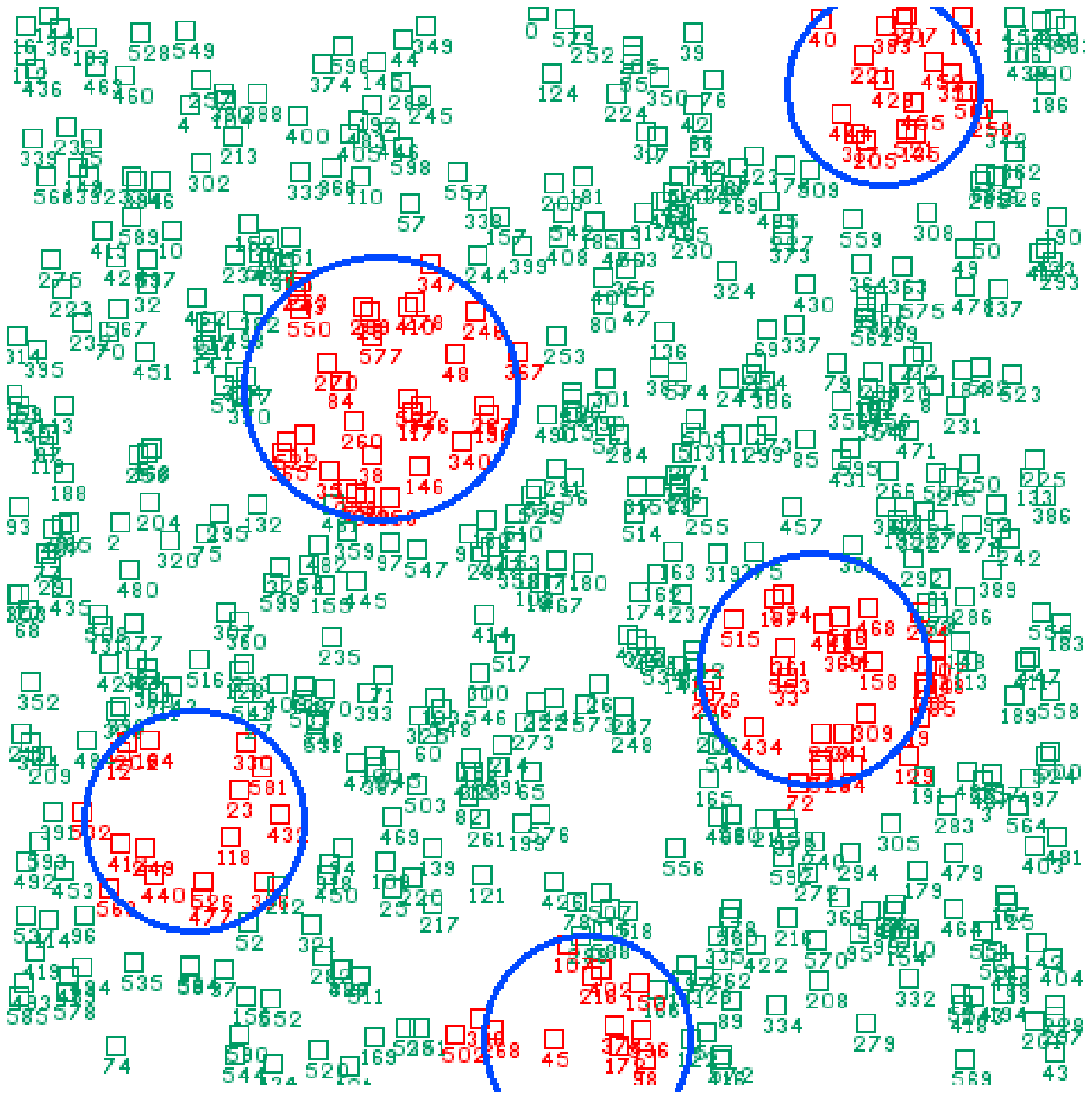}}}
  \end{center}
  \caption{Mapping sequence.}
  \label{fig:seq}
\end{figure}

In this section, we present a detailed description of our protocol for
mapping the jammed regions in a WSN. We used the sequence in
Fig.~\ref{fig:seq} as an example to guide our discussion. The mapping is
performed in few steps as described in following sections.

\begin{enumerate}
\item In the first step, the boundary nodes on the edge of a jammed
  region detects jamming in the network (\S~\ref{detect}), then
  some of them send notifications to the nodes outside the jamming
  range.

\item Some of the nodes that got notifications, report to the BS about
  the presence of jamming in the network.

\item The BS then locates the jammed regions and finally maps those
  regions based on the information it has received.
\end{enumerate}

\subsection{Jamming detection and notification to the BS}\label{detect}
A jamming device present in the network jams all the benign nodes within its signal range. We assume that the nodes are able to detect jamming of neighboring nodes. After detecting jamming some of these nodes send reports to the BS. We call the nodes that report jamming to the BS as {\em reporter} nodes. To keep the overall message overhead to a minimum, not all the nodes that detect jamming become reporter nodes. Whether a node will send a report to the BS depends on the following two decisions.
\begin{list}{\labelitemi}{\leftmargin=1em}
\item
{\bf Decision to become reporter:} When a node detects jamming among its neighbors, it decides with some probability $P_{rep}$, whether or not to become a reporter node. $P_{rep}$ should be determined according to the density of the network and also depending on the required accuracy of mapping. Once jamming starts and an unjammed node finds jamming among its neighbor nodes, it starts making a list of these jammed nodes. In Fig.~\ref{fig:rep}, the green nodes are the reporter nodes of the network with $P_{rep}$ = $0.5$, and the red ones are the jammed nodes that are reported to the BS.

\item
{\bf Decision to notify neighbors:} The farther a reporter node is located from the BS, the greater the number of messages required for sending jamming notifications to the BS. To reduce this overhead while ensuring good reporting coverage, we use a simple scheme where the reporter node multiplies its distance to the BS with the number of its neighbors and  $P_{rep}$. If this value is greater than a threshold value $T_{rep}$, the node will send an alert to its neighbors. If a node is the first one to notify its neighbors, it will send a report to the BS, otherwise it does not. If this value is less than or equal to $T_{rep}$, the node will send the report to the BS without sending any alerts to its neighbors. For example, a reporter node i will alert its neighbors if Eq.~\ref{th_eq} is fulfilled.
\begin{equation}
Dist_{i} \times Neighbor_{i} \times P_{rep} \leq T_{rep}
\label{th_eq}
\end{equation}
Here, $Neighbor_{i}$ is the number of neighbors of the reporter node and $Dist_{i}$ is its distance from the BS. 
A node that gets an alert from a neighboring reporter node will not send any notification messages to the BS, even if it has selected itself as a reporter node. 
\indent
The equation has been designed so that if the node is too far from the base station and has many neighbors the probablilty of it sending a message to the base station. The value $T_{rep}$ is determined based on the average distance to the BS and average number of neighbors for the sensor nodes in the network. There is a trade off between the cost of sending a notification from a reporter node to the BS and of sending notifications to the reporter's neighbors. $T_{rep}$ should be chosen such that it reduces the message overhead. 

\end{list}

\subsection{Locating the jammed regions of the network}
\label{reg_find}

Once the BS starts receiving notifications of jamming from the un-jammed nodes located at the various parts of the network, it starts building its own list of known jammed nodes in the network. Since the BS may get notifications of jamming from different parts of the network, it applies a clustering algorithm to decide on the number and location of the jammed regions in the network. For this purpose, we used k-means \cite{KMEANS, KMEANS67}, which is a partition-based clustering algorithm. The geographical positions of the nodes are used as data points and the euclidean distance between nodes is used as the distance measure. The output of the algorithm are clusters, which represent the jammed regions in the network.

$k$ is one of the inputs to the k-means clustering and in this case it stands for the number of jammed regions in the network. This value is not known apriori to the BS. If there is no manual assistance available to determine the probable value of $k$, the BS needs to decide on the best value of $k$, based on the information it receives from the nodes in the network.

Since, the quality of the clustering by k-means greatly depends on the selection of initial centroids, the BS runs the algorithm on a maximum value of $k$ for a certain number of times, each time using $k$ random points as initial centroids. So, the BS starts from $k = k_{max}$ and picks $k$ random nodes among its list of jammed nodes to serve as $k$ initial centroids. The BS then runs the k-means algorithm a fixed number of times for this value of $k$. In each round, the BS picks the best result with the lowest sum of squared error ($SSE$). For the next round, it merges two of the closest clusters from the previous best result and generates a new centroid from the mean of the centroids for these two clusters. The BS then uses this new centroid and the other $k-2$ centroids as initial centroids for the next round of k-means for $k-1$ clusters. Except for the first round with the maximum value of $k$, BS runs the k-means algorithm only once for each decrement of $k$, until $k=1$.

After generating $k_{max}$ number of clusters by the process described above, the BS decides on the optimal value of $k$ from the results. To do this, BS compares the improvements in the $SSE$ value for each of the clustering results from $k$ = $2$ to $k_{max}$ as improvement $Imp_{k}$ --
\begin{equation}
Imp_{k} = ({SSE_{k-1} - SSE_{k}}) / SSE_{k-1}
\end{equation}

\noindent If $Imp_{k}$ is negative for a given value of $k$, clustering results for that $k$ are discarded. This is because of the intuition that with the increase of $k$, $SSE$ should always decrease.

From the remaining clustering results, starting from the lowest value of $k$, the BS checks the nodes belonging to the cluster corresponding for that value of $k$ and accepts the clustering result if all the nodes are within two standard deviations from the mean of the cluster. If any of the nodes is more than two standard deviations away from the mean, the cluster is discarded. Otherwise, this $k$ is decided to be the number of jammed regions in the network. 

When the value of $k$ that is decided by BS is less than or greater than the original number of jammed regions in the network, one of the following two situations occurs - 
\begin{itemize}
\item
$k <$ number of jammed regions: If $k$ is less than the number of jammed regions in the network, multiple regions can be grouped into a single region. This causes, more un-jammed nodes to be inside a mapped region, thereby generating a larger FP rate. 
\item
$k >$ number of jammed regions: If $k$ is greater than the number of jammed regions in the network, a single region can be divided in to two or more regions. This will create multiple small regions in place of a single larger one, thus leaving some jammed nodes outside the mapped areas and causing a larger FN rate. 
\end{itemize}

\subsection{Jammed region mapping}

After the BS first decides on the number and location of the jammed regions present in the network, it maps the estimated jammed area for each region. By mapping the regions, the BS works as a classifier for a two-class prediction problem, where the outcomes are labeled as positive (p) or negative (n) class. The jammed nodes are labeled as positive and the un-jammed nodes are labeled as negative. If a node inside a mapped region is actually jammed then it is labeled as true positive (TP); however if the node inside the mapped region is un-jammed, then the outcome is a false positive (FP). Conversely, a true negative (TN) can occur when an un-jammed node falls outside the mapped regions and false negative (FN) happens when a jammed node falls outside the mapped regions. Also, the \emph{known positive} (TP/FP) nodes to the BS are the jammed nodes that were reported to BS by the reporter nodes and the \emph{known negative} (TN/FN) nodes to the BS are these reporter nodes. 

The mapping is performed based on the fact that jamming devices (using an omni directional antenna) usually create have circular shaped jamming regions in the network. The output of the mapping protocol is shown in Fig.~\ref{fig:area} which shows the original network with the mapping by the BS.

To map the region for each of the jammed areas found by BS, it first determines the center for that particular region, then fits the jammed nodes in a circular region and finally moves and increase the size of this region to increase the number of known TP and while keeping the number of known FP to minimum.

The steps are as follows-

\begin{list}{\labelitemi}{\leftmargin=1em}
\item
{\bf Determining the center for the region:} The BS first determines the center ($C_{j}$) for a region.


If the number of nodes in a cluster is less than $3$, the center is the mean of the points in the cluster.

Otherwise, the BS finds the convex hull of the cluster and the mean of the vertices of this hull is used as the center for that region.

The reason for finding the convex hull is to minimize the effect of the center ($C_{j}$) being biased to the side of the region with higher density. We use the Graham scan algorithm to find the convex hull. 

\item
{\bf Mapping the region:}
After finding the centroid ($C_{j}$) of the jammed nodes belonging to a region, the BS takes the largest distance between any two jammed nodes of this region as the diameter of a circle centered at $C_{j}$ and finds the number of known jammed nodes (TP) and reporter nodes (known FP) that are inside this circle. It also calculates the center ($C_{i}$) of the known FP nodes (Fig.~\ref{fig:areamap2}).

\item
{\bf Improvement in Mapping:}
To improve the mapping, the BS moves $C_{j}$ one step (away from $C_{i}$) at a time, alternating between moves along the vertical and horizontal axes. It continues to move the circle until either any known jammed node goes out of the circular region or if the number of known FP increases. If new TP nodes are added to the region, the BS increases the diameter of the circle by a factor to consume more jammed nodes. Fig.~\ref{fig:areamap2} shows the final area mapped by the BS with the new centroid for the jammed nodes located at ($C_{j}'$).

Fig.~\ref{fig:areamap2} shows the region mapped by the BS based on the information collected on the jammed nodes from the reporter nodes. Here, the orange nodes are true positives (TP), the green nodes are false positives (FP) and the red nodes are false negatives (FN) among the nodes found as jammed by the BS.

\begin{figure}
  \begin{center}
    \subfloat[Initial mapping]{\label{fig:initial}\fbox{\includegraphics[width = 1.5in]{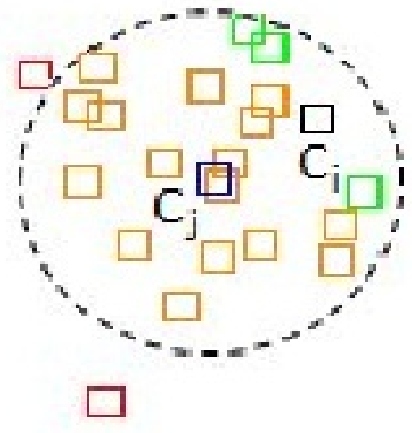}}}
    \hspace{.1in}    
    \subfloat[Final mapping]{\label{fig:final}\fbox{\includegraphics[width = 1.5in]{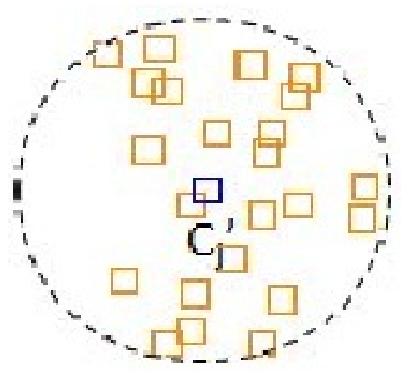}}}
  \end{center}
  \caption{Jammed region mapping by the BS.}
  \label{fig:areamap2}
\end{figure}

\end{list}

\noindent
In case of jamming regions that are asymmetric or non-circular shaped, the BS can use the result of the previously calculated convex hull to map the region.

\section{Simulation}
\label{sim}
We built a simple simulator to evaluate our proposed system. We use the results to evaluate the performance of our system and to compare our proposed system with JAM~\cite{JAM}. Our comparison is based on two metrics: 1) the time to perform the mapping of all the jammed regions present in the network, and 2) the amount of messages exchanged among the sensor nodes to do the mapping. 

Also, to evaluate the performance of our system, we measure the performance of the mapping done by the BS in terms of true positives, false positives, and false negatives (incrementing the node selection probability leads to minimizing the number of false negatives and false positives).  

In the following sections we describe the experimental setup, methods for comparison with JAM, and then the performance of the system.

\subsection{Experimental setup}
We built the simulator in C++ to simulate a border region with dimensions of $200 \times 200$ units. In this simulator, we study a WSN which is composed of between $500$ and $1000$ nodes. These nodes are randomly deployed in this area with one node near the upper horizontal border of the network serving as the BS (Fig.~\ref{fig:initial}) whom all the other nodes report. During the deployment phase of the network, for each node, the neighbor discovery and path setup to the BS is performed. The nodes have a fixed signal radius ($10$ -- $20$ units) and have $7$ -- $17$ neighbors on average.

During our simulation of events, an intruder places jamming devices randomly in the network. These devices have higher signal radii ($17$ -- $27$ units) than the sensor nodes, and this range can be different for the individual jamming devices in the network. When any of the nodes in the network is jammed, the neighboring nodes are notified by the detection process. When a working node discovers jamming among its neighbors, it decides to become a reporter node with probability $P_{rep}$.

\subsubsection{Comparison with JAM} 
We compare the performance of JAM \cite{JAM} and our proposed system. To do this we simulated both JAM and our system in the exact same network conditions.
\newline
\noindent {\bf Time to Map:} We measured the time to map all the jammed regions in the network by the BS. 

For JAM, the time to get the mapping is measured by the number of coalitions of jammed groups that occur at a mapping node. This number determines the number of rounds of build messages that are going to be sent by the mapping nodes before the final mapping result is being sent to BS. By this protocol, after the mapping is done, there should be ideally one dominant group of mapped (jammed) members and only one mapping node sends this information to BS.

For our system, total time to map is calculated by the number of alert messages a reporter node sends to its neighbors, which determines the time it takes for BS to receive all the notifications of jammed nodes in the network. \newline

\noindent {\bf Message overhead:}
The final goal is to have the BS learn about all the jammed regions in the network. To acheive this, during simulation, after mapping is done according to JAM, exactly one node (ideally the creator of the dominant group) from each jammed region sends a message to the BS containing information on the related mapped nodes. In our protocol, the overhead is the sum of the number of messages required to be sent by all the reporter nodes to notify the BS.

For the first set of experiments, we consider networks of three different densities -- $600$, $800$, and $1000$ nodes. For each of these settings, the simulation results are calculated by running JAM and our system with $P_{rep} = 0.2, 04, and 0.6$. For each of these cases the simulation has been run $100$ times and results are calculated by the arithmetic mean of these simulations. 

\begin{figure}
  \begin{center}
    \subfloat[]{\fbox{\includegraphics[width = 3in]{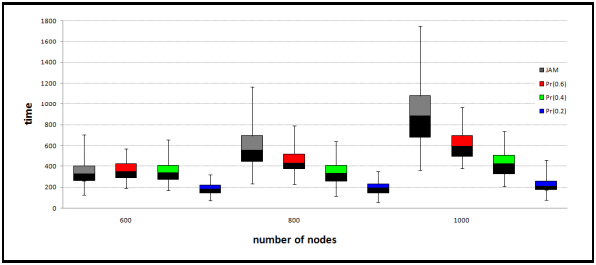}}}
  \end{center}
  \caption{Comparison with JAM on time to map jammed regions.}
  \label{fig:comp_time}
\end{figure}

\begin{figure}
  \begin{center}
    \subfloat[]{\fbox{\includegraphics[width = 2.25in]{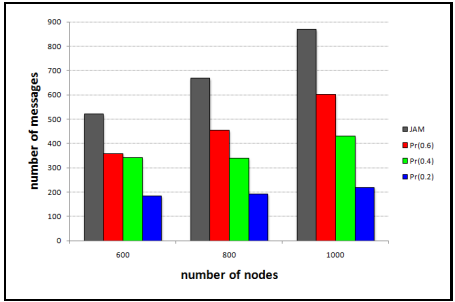}}}
  \end{center}
  \caption{Comparison with JAM on message overhead.}
  \label{fig:comp_msg}
\end{figure}

The box-and-whisker graph in Fig.~\ref{fig:comp_time} shows the time to get the ultimate mapping result for JAM and our system for three different probabilities and for three different densities of the network.
In the bar graph in Fig.~\ref{fig:comp_msg}, we compare the overhead of the two schemes according to the number of messages.

\subsection{Performance of the system}

In the second set of experiments, we evaluate the performance of our system. We took a measure of performance by incrementing the probability of a node to become reporter, varying $P_{rep}$ from $0.2$ to $0.6$. We present performance in terms of \emph{precision} and \emph{recall}. These two values are computed by the of number of true positive, false positive and false negative nodes the BS can identify while performing the mapping of the jammed nodes. 

Here, precision (Eq.~\ref{pre}) gives the probability of finding real jammed nodes classified as jammed by the BS to all then nodes classified as jammed by the BS. On the other hand recall (Eq.~\ref{rec}) specifies the probability of original jammed nodes mapped by the BS among all the jammed nodes present in the network. Here, TP, FP and FN stands for true positive, false positive and false negative values for the jammed nodes identified by the BS, respectively.

\begin{equation}
 Precision =\frac{TP}{TP+FP} 
\label{pre}
\end{equation} 

\begin{equation}
Recall =\frac{TP}{TP+FN}
\label{rec}
\end{equation} 

We did the experiment with three different densities of network of -- $600$, $800$ and $1000$ nodes. Results from the second set of experiments are shown in Table~\ref{precall}. Recall improves better with increasing probability as when the number of reporter nodes are higher. This is because the BS has more information about the jammed nodes, which helps the BS to include more jammed nodes in each region. The improvement in precision occurs in slower rate as even with the more information on the jammed nodes the BS can include some unjammed nodes in the clusters that are unknown to it.

\begin{table}[tp]
\begin{center}
\caption{Precision and recall}
\begin{tabular}{|c||c|c|c|} \hline
	& Probability & Precision & Recall \\ \hline\hline
	600 nodes &
	  \begin{tabular}{c} 0.2 \\ 0.4 \\ 0.6 \\ 
        \end{tabular} & 
      \begin{tabular}{c} 0.77 \\ 0.81 \\ 0.82 \\  
        \end{tabular} &
      \begin{tabular}{c} 0.51 \\ 0.73\\ 0.75 \\ 	
        \end{tabular} \\ \hline\hline
	800 nodes &
	  \begin{tabular}{c} 0.2 \\ 0.4 \\ 0.6 \\ 
		\end{tabular} &
      \begin{tabular}{c} 0.79 \\ 0.83 \\ 0.84 \\
        \end{tabular}  &
      \begin{tabular}{c} 0.56 \\ 0.73 \\ 0.80 \\
        \end{tabular} \\ \hline\hline
	1000 nodes &
	  \begin{tabular}{c} 0.2 \\ 0.4 \\ 0.6 \\
		\end{tabular} &
      \begin{tabular}{c} 0.83 \\ 0.84 \\ 0.87 \\ 
        \end{tabular}  &
      \begin{tabular}{c} 0.59 \\ 0.75 \\ 0.82 \\ 
        \end{tabular} \\ \hline		  
\end{tabular}
\label{precall}
\end{center}
\end{table}


\section{Experiment}
\label{exp}

In this section we describe a set of experiments that we conduct using real sensor motes to test the performance of the jammed region mapping technique. We use Crossbow's TelosB motes (TPR2400), which are generally platform for low-power research development for wireless sensor network experiments and TinyOS 2.0 for programming.

\subsection{Setup}

For our experiments we use a total of $50$ motes, of which $49$ are used for the network, and one is programmed as jamming device.
We place the motes in a $7 \times 7$ grid. We placed adjacent motes $8''$ apart. Table~\ref{exp_env} summarises the network settings.

The network area is a square of size $48'' \times 48''$. Since the default radio range of the sensor motes is quite high, We set the parameter DCC2420-DEF-RFPOWER of the motes to $1$, $2$ and $3$ and found $9''$, $20''$ and $213''$ as corresponding ranges. For our indoor experiments, we used the lowest range (DCC2420-DEF-RFPOWER = $1$).

\begin{table}
\begin{center}
\caption{Basic setup of the network}
\begin{tabular}{ | l | p{1.5cm}|}
\hline
Size of Network & 49\\ \hline 
Average number of neighbors & 5.22\\ \hline
Total number of jammed nodes & 10\\ \hline
Sensor node signal range & $9''$\\ \hline
Jammer signal range & $9''$\\ \hline
Average number of jammed regions & 1\\ \hline
\end{tabular}
\label{exp_env}
\end{center}
\end{table}

\subsubsection{Jammer node}
To make a regular sensor node to work as the jammer, we bypass the MAC protocol for that mote by disabling channel sensing and radio back off operation. This allows the jammer mote to send continuous signals and jam reception of all the motes those are within its transmission range. 

\subsubsection{Neighbor setup}
Each sensor node in the network sends a beacon packet to its neighbors at regular time intervals ($0.4s$). We determine the neighbors of a node by setting a threshold for the number of messages it receives from another node at unit time.

\subsubsection{Detection}
The beacon packets sent by the motes to its neighbors are used to detect jamming in the network. When the jammer node is on, the jammed sensors fail to receive the regular beacon packets from its neighbors and will be able to detect that its being jammed. The boundary nodes of the jammed region notify their un-jammed neighbors that they are jammed and the un-jammed nodes starts the mapping protocol. 

\subsection{Simulation}
We give this network setup with each node and its corresponding neighbors and also the list of jammed nodes from the experiment as input to our simulator (section~\ref{sim}) and see how our region mapping protocol performs for actual jammed regions. Since the jammed regions that are generated are of irregular sizes, we used a convex-hull finding algorithm for the final mapping, instead of the circular shapes. For each experiment, we run our simulation of region mapping for 1000 times to measure the performance, by selecting random reporter nodes for each round.

The performance in terms of precision and recall is presented in Table.~\ref{exp_result}.

\begin{table}
\begin{center}
\caption{Results from simulation of the network}
\begin{tabular}{ | l | p{1.5cm}| p{1.5cm} | p{1.5cm}|}
\hline
Pr. of selection & Precision & Recall\\ \hline\hline
0.4 & 0.50 & 0.25\\ \hline
0.6 & 0.81 & 0.54\\ \hline
0.8 & 0.94 & 0.78\\ \hline
\end{tabular}
\label{exp_result}
\end{center}
\end{table}

\section{Conclusion}
\label{conc}

Jamming is a critical attack against WSNs, which can lead to a DoS condition. This can prevent the network from monitoring for intruders. Also jamming can be used against other application scenarios. Also, an attacker can jam random parts in a network and create a path for himself to go back and forth through the network, thus creating a critical security breach. It is important to map out the jammed regions so that this information can be used in the network for routing and power management and also for taking reactive measures to deal with unmonitored regions. Thus, jammed region mapping in the network helps dealing with jamming and to take effective measures against it. 
\par In this paper, we proposed an efficient mapping protocol to map the jammed regions in a network by having the base station compute and approximate mapping. This relieves the sensor nodes from sending many mapping messages and running out of the battery power. The mapping results can be improved by having more nodes send jamming notification messages to base station, which creates a trade off between performance of mapping versus the network overhead. Our simulation results demonstrate that this system requires less interactions among the sensor nodes compared with previous work and thus has less overhead and faster mapping. 
\par We developed our intruder model assuming jamming devices placed randomly in the network and creating circular interference patterns. In future work, we will study mapping of jamming regions introduced by jamming devices of more asymmetric and irregular signal range. We will investigate the application of improved k-means algorithm~\cite{KMEANSIMP1, KMEANSIMP2}, so that we can have better selection of the initial centroids and improved clustering results in the presence of clusters of irregular size, density, and shape.




\balance
\bibliographystyle{plain}
\bibliography{bibliography}

\balance
\end{document}